\documentclass[final,5p,times,twocolumn]{elsarticle}

\usepackage{graphicx}
\usepackage{color}

\usepackage{amssymb}

\usepackage[nodots]{numcompress}

\usepackage{lineno}

\journal{Current Opinion: Colloid and Interface Science}

\RequirePackage[normalem]{ulem} 
\RequirePackage{color}\definecolor{RED}{rgb}{1,0,0}\definecolor{BLUE}{rgb}{0,0,1}

\begin{document}

\begin{frontmatter}



\title{Active colloids in complex fluids}


\author{Alison E. Patteson$^{1,*}$, Arvind Gopinath$^{2,*}$ \& Paulo E. Arratia$^{1,**}$}

\address{$^{1}$Department of Mechanical Engineering \& Applied Mechanics, University of Pennsylvania, PA 19104.}
\address{$^{2}$School of Engineering, University of California Merced, CA 95340.}
\address{$^{*}$Equal contribution \\
$^{**}$Corresponding Author \\ Email address: parratia@seas.upenn.edu}

\begin{abstract}

We review recent work on active colloids or swimmers, such as self-propelled microorganisms, phoretic colloidal particles, and artificial micro-robotic systems, moving in fluid-like environments. These environments can be water-like and Newtonian but can frequently contain macromolecules, flexible polymers, soft cells, or hard particles, which impart complex, nonlinear rheological features to the fluid. While significant progress has been made on understanding how active colloids move and interact in Newtonian fluids, little is known on how active colloids behave in complex and non-Newtonian fluids. An emerging literature is starting to show how fluid rheology can dramatically change the gaits and speeds of individual swimmers. Simultaneously, a moving swimmer induces time dependent, three dimensional fluid flows, that can modify the medium (fluid) rheological properties. This two-way, non-linear coupling at microscopic scales has profound implications at meso- and macro-scales: steady state suspension properties, emergent collective behavior, and transport of passive tracer particles. Recent exciting theoretical results and current debate on quantifying these complex active fluids highlight the need for conceptually simple experiments to guide our understanding.

\end{abstract}


\end{frontmatter}


\section{Introduction}

Active fluids are ubiquitous in nature and permeate an impressive range of length scales, ranging from collectively swimming schools of fish ($\sim$ km)~\cite{Vicsek2012} and motile ants ($\sim $mm) ~\cite{Goldman2015} to microorganisms ($\sim $ $\mu$m)~\cite{Berg2008,Lauga2009,Goldstein2015,Stocker2015} and molecular motors within individual cells ($ \sim$ nm) \cite{Howard2001,Nikta2014}. Suspensions of {active particles, commonly defined as self-propelling particles that inject energy, generate mechanical stresses, and create flows within the fluid medium}, constitute so-called {\em active} fluids ~\cite{Ramaswamy2010,Marchetti2013}. This internally-injected energy drives the fluid out of equilibrium (even in the absence of external forcing) and can lead to swirling collective behavior \cite{Kessler2004} and beautiful pattern formation \cite{BenJacob1997,Surrey2001}, that naively appear unique to life. Indeed, the motility of swimming microorganisms such as nematodes, bacteria, protozoa and algae has been a source of wonder for centuries now. Anton van Leeuwenhoek, upon discovering bacteria in 1676,  observed, ``\emph{I must say, for my part, that no more pleasant sight has ever yet come before my eye than these many thousands of living creatures, seen all alive in a little drop of water, moving among one another}'' \cite{dobell1932}. Since then, scientists have observed and classified other collective large-scale patterns in active fluids, such as vortices~\cite{Zhang2010,Vicsek1996}, flocks~\cite{Toner1998}, and plumes~\cite{Kessler1985,Kessler1995,Goldstein2004} that form at high concentrations of their organisms and highlight the link between life, fluid motion and complex behavior. Surprisingly, synthetic materials/particles have been recently developed which also exhibit these life-like complex behaviors. Examples are shaken granular matter ~\cite{Chate2010,Menon2007}, phoretic colloidal particles \cite{Bartolo2013,Palacci2013}, soft field-responsive gels \cite{Alexeev2012}, and included in this review externally-actuated artificial swimmers \cite{Dreyfus2005,Nelson2013,Wang2014, Peer2014}.

These active particles (living or synthetic, hard or soft), as collected in {Fig. 1}, have sizes that range from a few tenths of a micron to a few hundred microns, spanning colloidal length scales over which thermal noise is important~\cite{Aranson2013}. The motion of these active colloids allows one to either direct (channel) or extract (harness) the energy injected at one length scale at other scales. For instance, activity can render large, normally athermal spheres diffusive \cite{Wu2000} and yield controllable, directed motility of micro-gears \cite{Wong2013,Aranson2009_2,Leonardo2009}.

\begin{figure*}
\centering\includegraphics[width=0.8\linewidth]{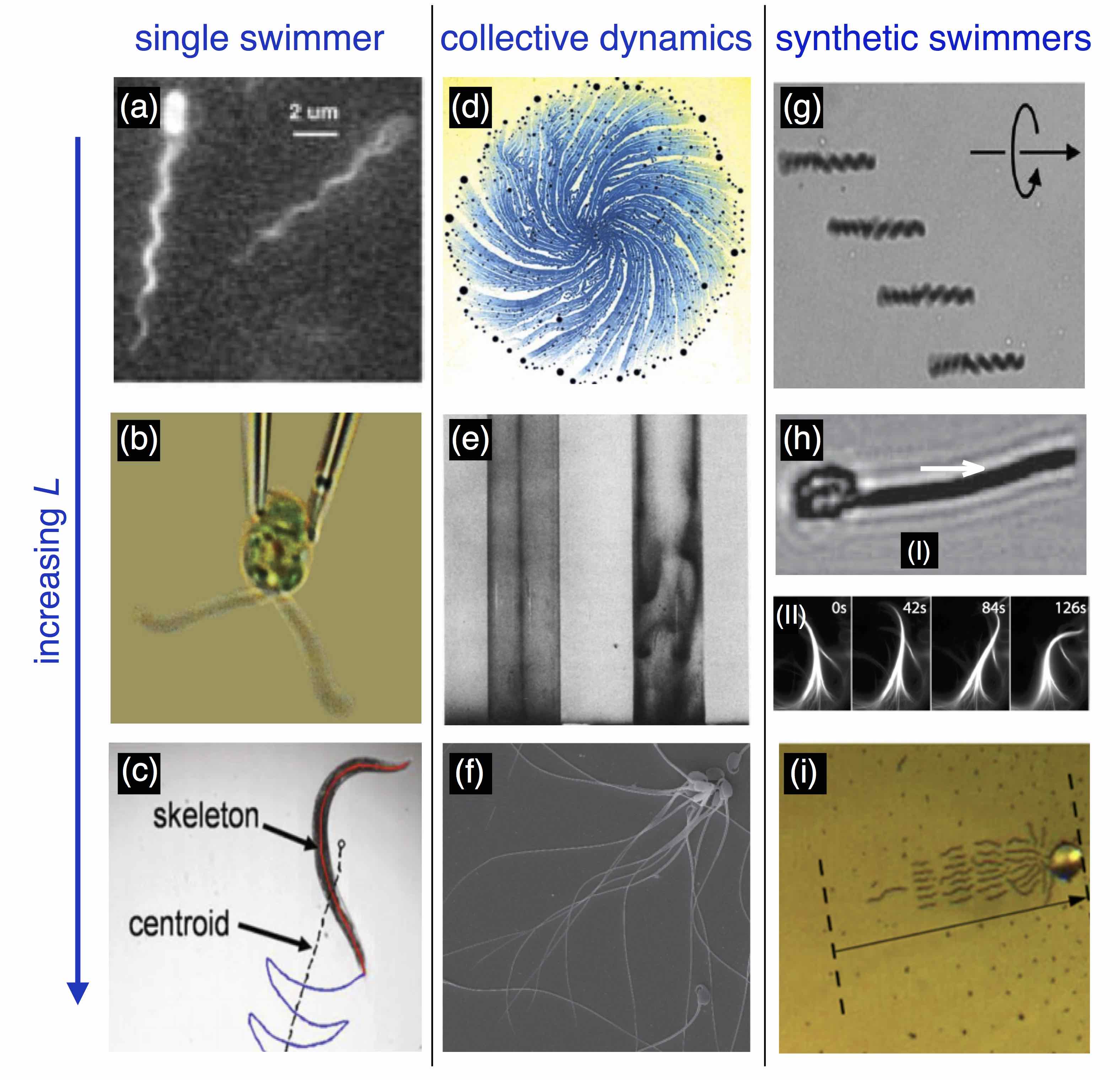}
\caption{An overview of active colloidal systems - natural and synthetic. (a)-(c): Individual natural swimming microorganisms arranged in order of increasing size:  (a) prokaryotic bacterium {\it Escherichia coli} with cell body approximately 2 $\mu$m \cite{Berg2000}, (b)  Eukaryotic unicellular alga {\it Chlamydomonas reinhardtii} with a cell body that is approximately 8 $\mu$m \cite{Goldstein2015}, and (c) multi-cellular organism {\it C. elegans} that is appoximately 1 mm long \cite{Sznitman2010}. (d)-(f) Examples of collective behavior seen in aggregates of microorganisms: (d) a bacterial colony of  {\it P. vortex} on agar \cite{Levine2001}, (e) bioconvection of algae under shear  \cite{Kessler1985}, and (f) cooperative behavior in sperm \cite{Maha2014}. (g)-(i) Synthetic swimmers: (g) field driven translation of helical magnetic robots \cite{Nelson2013_2}, (h:A) magnetically driven chain comprised of paramagnetic spheres attached via DNA strands \cite{Dreyfus2005}, (h:B) metachronal waves generated by reconstituted microtubule-motor extracts \cite{Dogic2011}, and (i) magnetically driven surface snakes comprised of self-assembled 80-100 $\mu$m spheres \cite{Aranson2009}.
}
\end{figure*}

Recently, there has been much interest in the production and dynamics of suspensions of active colloids \cite{Marchetti2013}. The study of such active suspensions is driven by both practical and scientific relevance. From a technological and engineering standpoint, active suspensions play an integral role in medical, industrial, and geophysical settings. The spread and control of microbial infections \cite{Josenhans2002,Costerton1999}, design of microrobots for drug delivery \cite{Gao2014} or non-invasive surgery \cite{Nelson2013}, biofouling of water-treatment systems \cite{Bhushan2012} and biodegradation of environmental pollutants \cite{Kessler2011} are just a handful of examples. From a scientific standpoint, active suspensions are interesting in their own right because they are non-equilibrium systems that exhibit novel and unique features such as turbulence-like flow in the absence of inertia \cite{Kessler2007,Yeomans2012}, anomalous shear viscosities \cite{Sokolov2009_2, Rafai2010,Clement2015}, enhanced fluid mixing \cite{Berg2004,Gollub2011}, giant density fluctuations \cite{Zhang2010,Menon2007} and liquid crystal like orientational ordering~\cite{Baskaran2012}. Because these features are generic to many other active materials (e.g. cell, tissues, vibrated granular matter), active suspensions can also serve as a toolbox for understanding and deciphering generic features of active materials across many length scales.
 
The suspending fluid in these active (colloidal) suspensions can be simple and Newtonian (e.g. water) or complex and non-Newtonian. Complex fluids are materials that are usually homogeneous at the macroscopic scale and disordered at the microscopic scale, but possess structure at intermediate scale. Examples include polymeric solutions, dense particle suspensions, foams, and emulsions. These complex fluids often exhibit non-Newtonian fluid properties under an applied deformation (e.g. shear) including viscoelasticity, yield-stress, and shear-thinning viscosity. An overarching goal in the study of complex fluids is to understand the connection between the structure and dynamics of the fluid microstructure to its bulk flow behavior \cite{Brady1988,Squires2009}. For example, recent experiments by Keim and Arratia \cite{Keim2013,Keim2014}, which visualize a monolayer of dense colloidal particles under cyclic shear at low strains, have shown how local particle re-arrangements connect to the suspension bulk yielding transition. This work highlights how local measures of the microstructure can shed new light on the bulk material response in an amorphous material.

In active fluids, it is even more challenging to link the activity at the microscale to the fluid meso- and macro-scales. For instance, living tissues are continuously exposed to stimuli, which can lead to growth and remodeling of their structure. This remodeling in the tissue microstructure is often implicated in medical conditions such as asthma. Recent work by Park {\it et al.} \cite{Lisa2015} has shown how tissue microstructural details, such as cell shape, affects bulk properties, such as fluidity and rigidity. In a similar vein, recent experiments have shown that the interplay between the motion of active particles and the complex fluid rheology of the suspending medium leads to a number of intricate and often unexpected results. In particular, the local mechanical stresses exerted by microorganisms in an active colloidal suspension can alter the local properties of its environment \cite{Wilson2014,Patteson2015}; while simultaneously, the complex fluid rheology modifies the swimming gaits and spread of individual organisms \cite{Shen2011,Boyang2015}. It is essential to understand this two-way coupling in order to uncover the universal principles underlying these active complex materials and in order to design and engineer new active materials.

In this paper, we review recent work on active colloids moving in fluidic environments and show how recent theory and experiments can elucidate the connections between microscale descriptions and the resulting macroscale collective response. We begin in Section 2 at the level of individually swimming colloids and how their motion couples to the suspending Newtonian (2.1) and non-Newtonian fluids (2.2). In Section 3, we focus on suspensions of swimming colloids at both dilute (3.1) and non-dilute (3.2) concentrations. A number of studies have shown that active colloids moving in Newtonian fluids (sections 3.1.1 and 3.2.1) can introduce non-Newtonian features, such as shear dependent bulk viscosities and viscoelasticity, to the suspension. In contrast, the change in the bulk rheology due to activity -- when the suspending fluid is itself non-Newtonian (sections 3.1.2 and 3.2.2) -- is just beginning to be explored theoretically as well as experimentally. {Figure 2 illustrates the importance of non-Newtonian fluid properties for individually swimming bacteria and for non-dilute concentrations of bacteria.} We conclude by highlighting new experimental techniques that will help address challenges and answer emerging questions related to active colloids in complex fluids.

\section{Fluid rheology and single swimmers}

\subsection{Single swimmer in Newtonian fluids}

Many organisms move in the realm of low Reynolds number Re $\equiv  \ell U \rho/\mu \ll 1$ because of either small length scales $\ell$, low swimming speeds $U$ or both. In a Newtonian fluid with density $\rho$ and viscosity $\mu$, this implies that inertial effects are negligible, the hydrodynamics is governed by the Stokes' equation, and stresses felt by the swimmer are linear in the viscosity. To therefore achieve any net motion (i.e. swim), microorganisms must execute non-reversible, asymmetric strokes as shown in {Fig. 3} in order to break free of the constraints imposed by the so-called ``scallop theorem"~\cite{Purcell1977}.

In the Stokes' limit, the flow caused by the moving particle can then be described as linear superposition of fundamental solutions such as stokelets and stresslets. The exact form of the generated flow depends on the type of swimmer. For instance, an externally-actuated swimmer with fixed gaits creates flow  that decay a distance $r$ away from the swimmer as $1/r$. A freely propelled swimmer is however both force free and torque free; therefore the induced fields are due to force dipoles, which decay as $1/r^2$, or higher order multipoles. Naturally-occurring, freely-moving organisms can typically be classified into one of two categories: (i) pullers  (negative force dipole) such as {\it Chlamydomonas reinhardtii} \cite{Goldstein2015} or (ii) pushers (positive force dipole) such as the bacteria {\it Escherichia coli} \cite{Berg2004} and {\it Bacillus subtilis} \cite{Zhang2010,Sokolov2009_2}. Note that other organisms such as the alga \textit{Volvox carteri} may fall between this pusher/puller distinction; other organisms move by exerting tangential waves along their surfaces and are called squirmers \cite{Lauga2009}. While this pusher/puller classification is limited and oversimplified, it provides a dichotomy for a reasonable framework.

\begin{figure*}
\centering\includegraphics[width=0.6\linewidth]{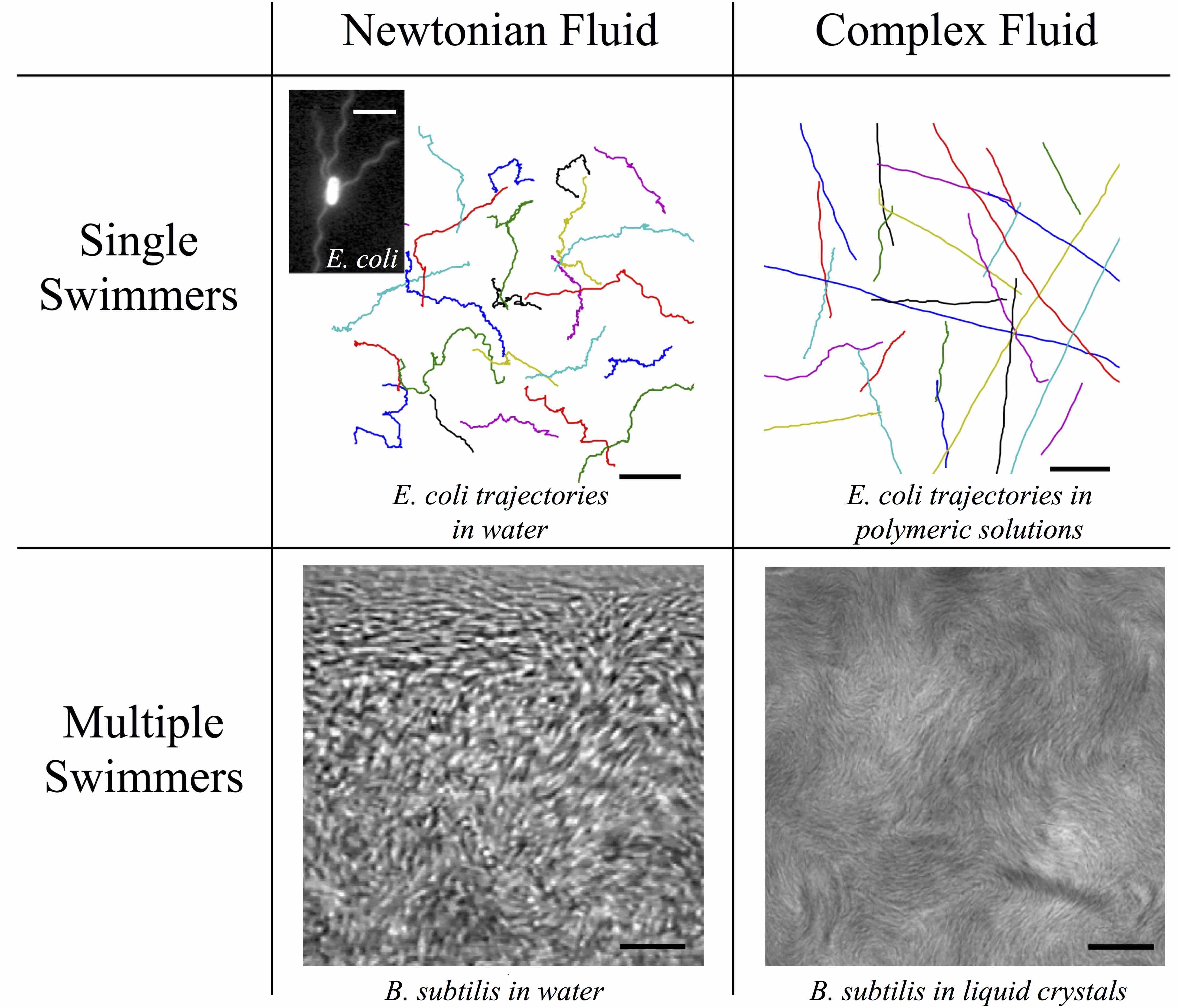}
\caption{{Swimming bacteria at dilute and non-dilute concentrations in Newtonian and non-Newtonian fluids. (a) {\it Escherichia coli} trajectories in water-like Newtonian fluid. Trajectories consist of straight segments (runs) punctuated by re-orienting tumbles. Scale bar is 15 $\mu$m. (Inset) A fluorescently-stained {\it E. coli} \cite{Berg2000}. Scale bar is 2 $\mu$m. (b) Replacing the Newtonian fluid with a polymeric solution results in straighter trajectories with suppressed tumbling and cell body wobbles \cite{Patteson2015}. Scale bar is 15 $\mu$m. (c) In Newtonian fluids, high concentrations of bacteria exhibit collective motion near an air/water interface \cite{Kessler2004}. Scale bar is approximately 15 $\mu$m. (d) In a liquid crystal, the swimming bacteria tend to align with the local nematic director. At high enough bacterial concentrations, the flow generated by the bacteria affects the long-range nematic order of the liquid crystal and creates dynamic patterns  of the director
and bacterial orientations
\cite{Aranson2013_2}. Scale bar is 30 $\mu$m.}}
\end{figure*}

\begin{figure*}
\centering\includegraphics[width=0.75\linewidth]{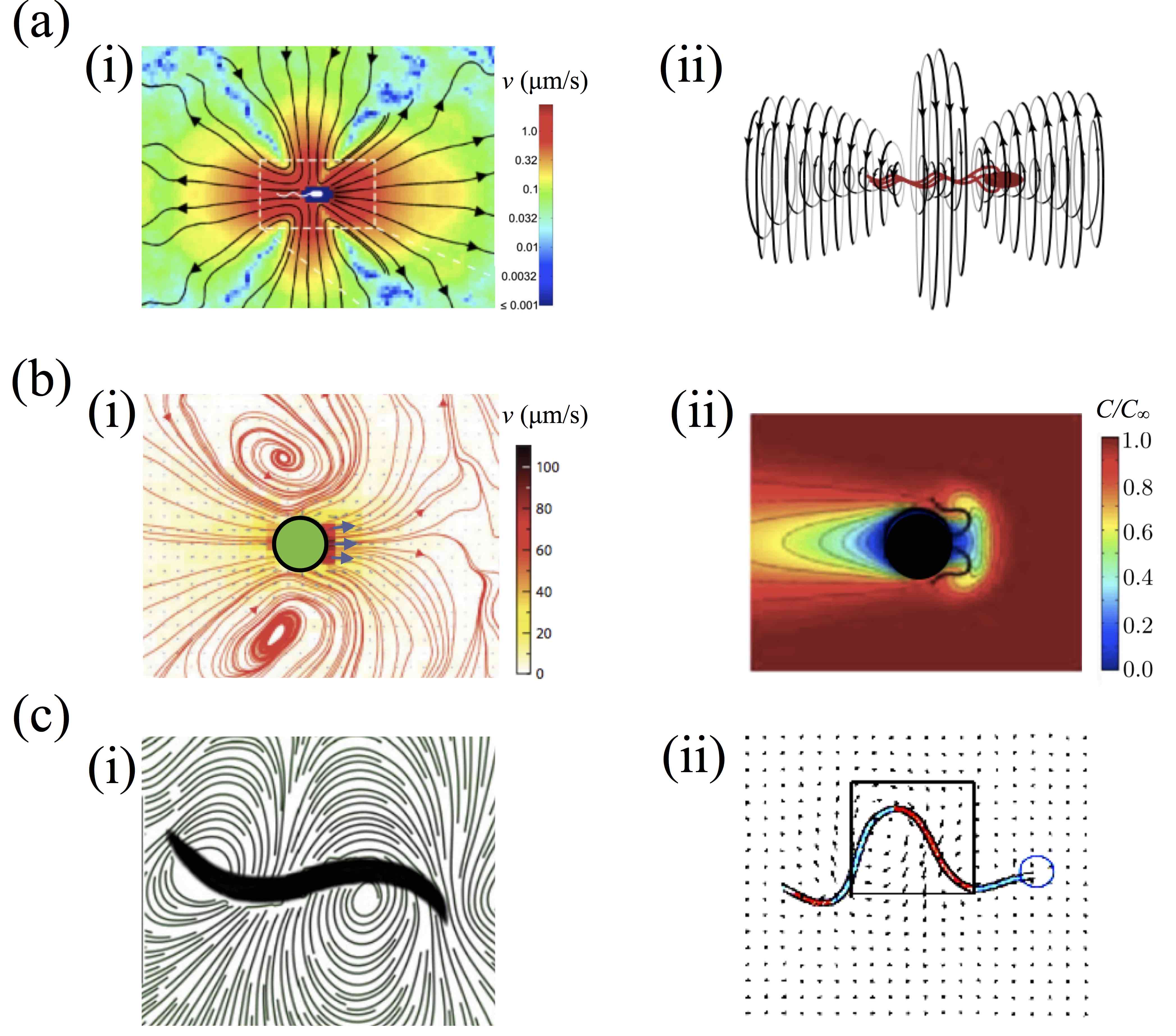}
\caption{Single natural swimmers moving in Newtonian fluids. (a) (i) Experimentally measured period averaged, color-coded velocity field around {\it Escherichia coli} bacterium \cite{Drescher2010}. (ii) Three-dimensional streamlines of a simulation of the flow in a frame co-moving with the bacterium \cite{Larson2010}. (b) (i) Averaged streamlines around {\it Chlamydomonas reinhardtii} \cite{Drescher2010_2}  - the color map denoting velocity magnitudes. (ii) Snapshots of the computed nutrient concentration fields $C$ around a model swimmer swimming in a nutrient gradient (B) \cite{Peko2010}. (c) (i) Streamlines around a swimming nematode {\it C. elegans} \cite{Gagnon2014_2}. (ii) Computed velocity fields around a flexible self-propelling swimmer \cite{Fauci2006}. }
\end{figure*}

The dipole approximations are useful in estimating force disturbances far from the swimmer. Closer to the moving swimmer, the flow field is time-dependent and can significantly deviate from these dipole approximations \cite{Gollub2010,Larson2010}. Significant theoretical work exists on characterizing these complex temporal and spatial flow fields around individual swimmers and obtaining approximate descriptions that may then be used as a first step in understanding how two and more swimmers interact \cite{Drescher2010}. Other geometries such as infinitely long waving sheets and cylinders have also been used to gain insight into the motility behavior of undulatory swimmers such as sperm cells and nematode (\textit{C. elegans}) \cite{Taylor1951,Lauga2014,Peko2010}. 

A feature common to these theoretical studies is that the swimming gait - i.e, the temporal sequence of shapes generating the propulsion - is assumed to be constant and independent of the fluid properties. Recent experiments paint a more colorful picture. Even in simple Newtonian fluids, fluid viscous stresses can significantly affect the microorganisms swimming gait and therefore their swimming speed \cite{Patteson2015,Boyang2015}.

 \subsection{Single swimmer in complex fluids}

The two-way coupling between swimmer kinematics and fluid rheological properties can give rise to many unexpected behaviors for microorganism swimming in complex fluids. For instance, the stresses in a viscoelastic fluid are both viscous and elastic, and therefore time dependent. Consequently, kinematic reversibility can break down and propulsion is possible even for reciprocal swimmers \cite{Keim2012, Gagnon2014}. This effect is especially important for small organisms since the time for the elastic stress to relax is often comparable to the swimming period \cite{Boyang2015,Guy2014}. Therefore, elastic stresses may persist between cyclic strokes. 

\begin{figure*}
\centering\includegraphics[width=0.7\linewidth]{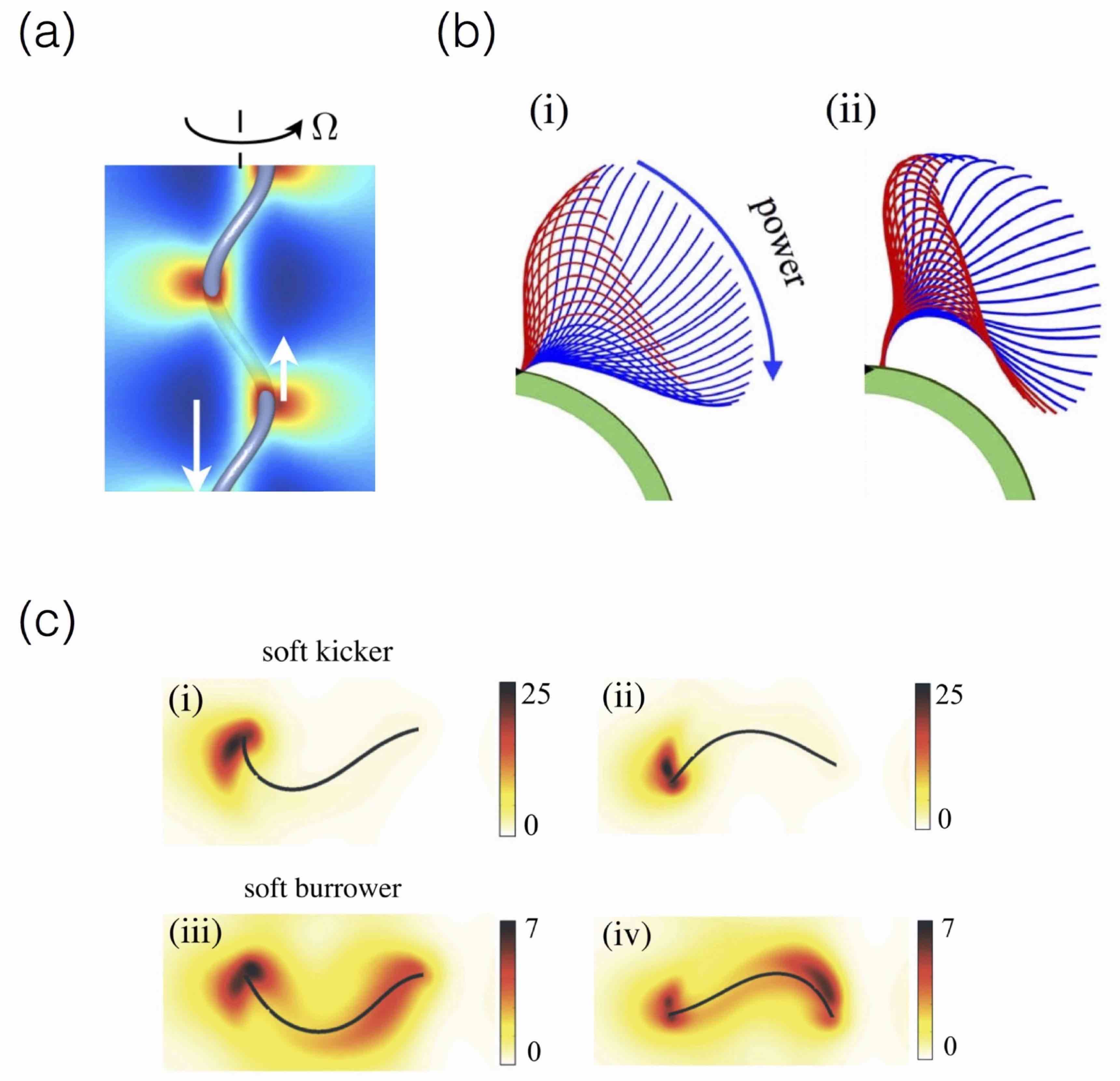}
\caption{ Single swimmers moving in viscoelastic fluids. (a) The axial component of fluid velocity generated by a rotating, force-free helical segment \cite{SS2013}. (c) (i) The sequence of shapes (swimming gait) 
attained by the cilia in {\it Chlamydomonas reinhardtii} in Newtonian fluid of viscosity around 6 Pa.s. The direction of the power stroke is indicated. (ii) The ciliary shapes seen when the same organism moves in a viscoelastic fluid are dramatically different \cite{Boyang2015}. (d) Contour plots of the polymers stress generated around a moving soft swimmer for a (i,ii) soft kicker and a (iii, iv) soft burrower. The mobility is affected by both the softness of the swimmer as well as by the elasticity of the fluid through which the swimmer moves \cite{Guy2014}.}
\end{figure*}

Emerging studies - some of which are highlighted in {Fig. 4} - are revealing the importance of fluid rheology on the swimming dynamics of microorganisms. Consider the effects of fluid elasticity on swimming at low Re. Would fluid elasticity enhance or hinder self-propulsion? Theories on the small amplitude swimming of infinitely long wave-like sheets \cite{Lauga2007} and cylinders \cite{Powers2007} suggest that fluid elasticity can reduce swimming speed, and these predictions are consistent with experimental observations of undulatory swimming in {\it C. elegans} \cite{Shen2011}. On the other hand, simulations of finite-sized moving filaments \cite{Shelley2010} or large amplitude undulations \cite{Guy2014} suggest that fluid elasticity can increase the propulsion speed - consistent with experiments on rotating rigid mechanical helices \cite{Liu2011}. In recent work on {\it Chlamydomonas reinhardtii}~\cite{Boyang2015}, we found that the beating frequency and the wave speed characterizing the cyclical bending of the flagella are both enhanced by fluid elasticity. Despite these enhancements, the net swimming speed of the alga is hindered for fluids that are sufficiently elastic. In complementary studies on {\it E. coli}, we found that the swimming velocity of the bacteria is enhanced for fluids that are sufficiently elastic \cite{Patteson2015}. Visualization of individual fluorescently labeled DNA polymers reveals that the flow generated by individual {\it E. coli} is sufficiently strong to stretch polymer molecules, inducing local elastic stresses in the fluid. These elastic stresses suppress inefficient wobbling, acting on the {\it E. coli} body in such a way that the cells swim faster. Overall, the emerging hypothesis is that there is no universal answer to whether motility is enhanced or hindered by viscoelasticity or shear-thinning viscosity. Instead, the microorganism propulsion speed in complex fluids depends on how the fluid microstructure (e.g. polymers, particles) interact with the velocity fields generated by a microorganism.

\begin{figure*}
\centering\includegraphics[width=\linewidth]{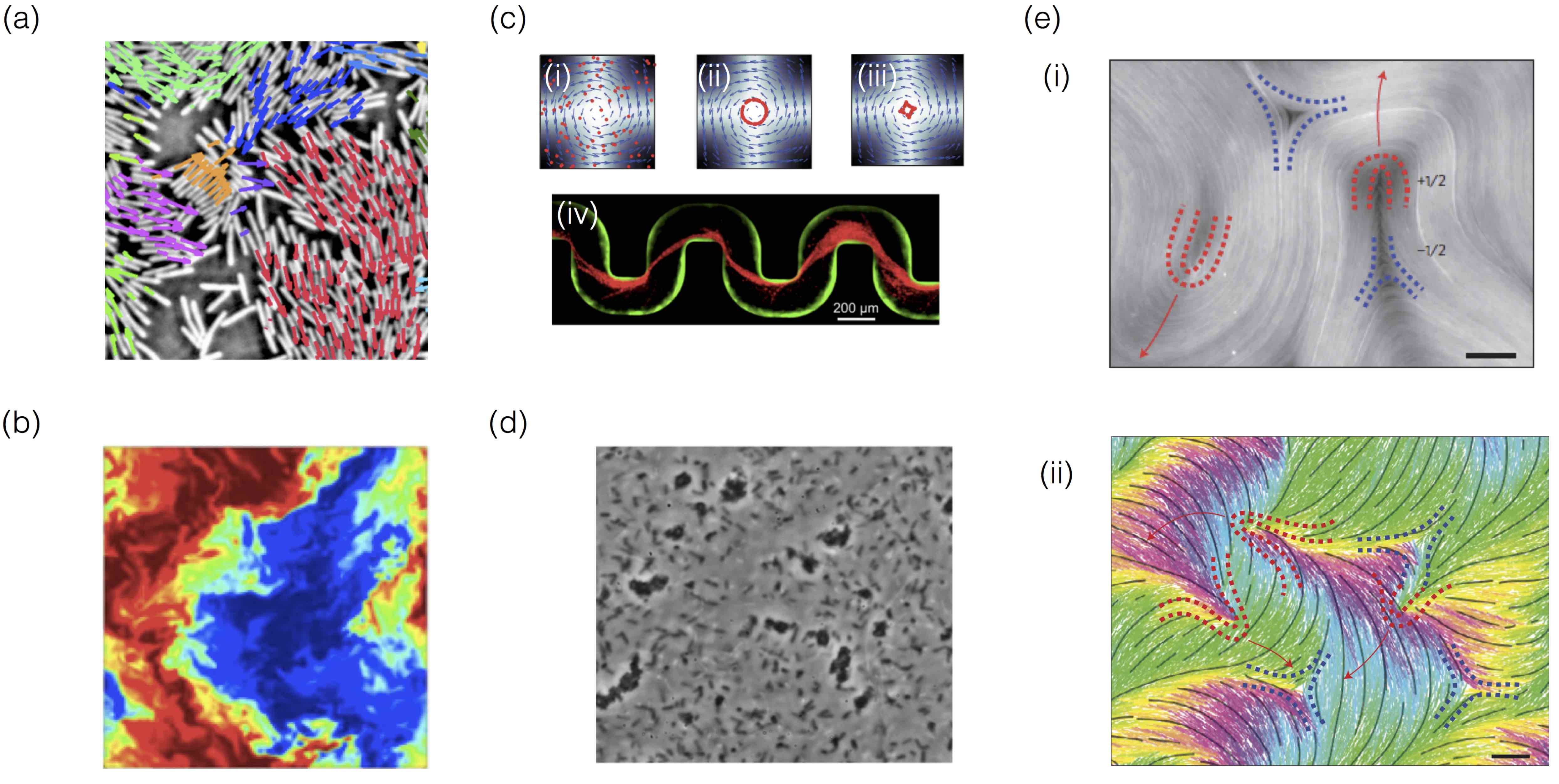}
\caption{Collective dynamics in active colloids. (a) A snapshot of a swarming bacterial colony of {\it B. subtilus} on agar. Velocity vectors are overlayed on the bacteria. The lengths correspond to the speeds and are used to identify individual clusters \cite{Zhang2010}. (b) Scalar fields such as tracer concentration can be passively advected by the background velocity field generated by motile bacteria,  demonstrating the mixing efficiency of active suspensions \cite{Shelley2011}. (c) (i-iii) Simulated spatial distribution of microorganisms in Taylor-Green vortices for different mobilities and elasticities. Relatively higher elastic effects cause an initially uniform distribution of microorganisms to aggregate. In real systems, where bacteria secrete polymers, this effect may enhance the aggregation and biofilm formation \cite{Arkedani2012}.(iv) Bacterial biofilm streamers (red) form efficiently at high bacterial concentrations and may lead to catastrophic blockage in synthetic and natural channels through which fluids flow \cite{Stone2013}. (d) Bright field microscopy image of a mixture of motile bacteria and polymers, evidencing the formation of bacterial clusters due to depletion effects \cite{Wilson2012}. (e) (i) Florescence microscopy image of a microtubule active nematic with defects of charge +1/2 (red) and -1/2 (blue). (ii) Snapshot of simulated nematic with marked defects. The color of the rod indicates its orientation and the black streamlines guide the eye over the coarse-grained nematic field \cite{Decamp2015}.}
\end{figure*}

\begin{figure*}
\centering\includegraphics[width=\linewidth]{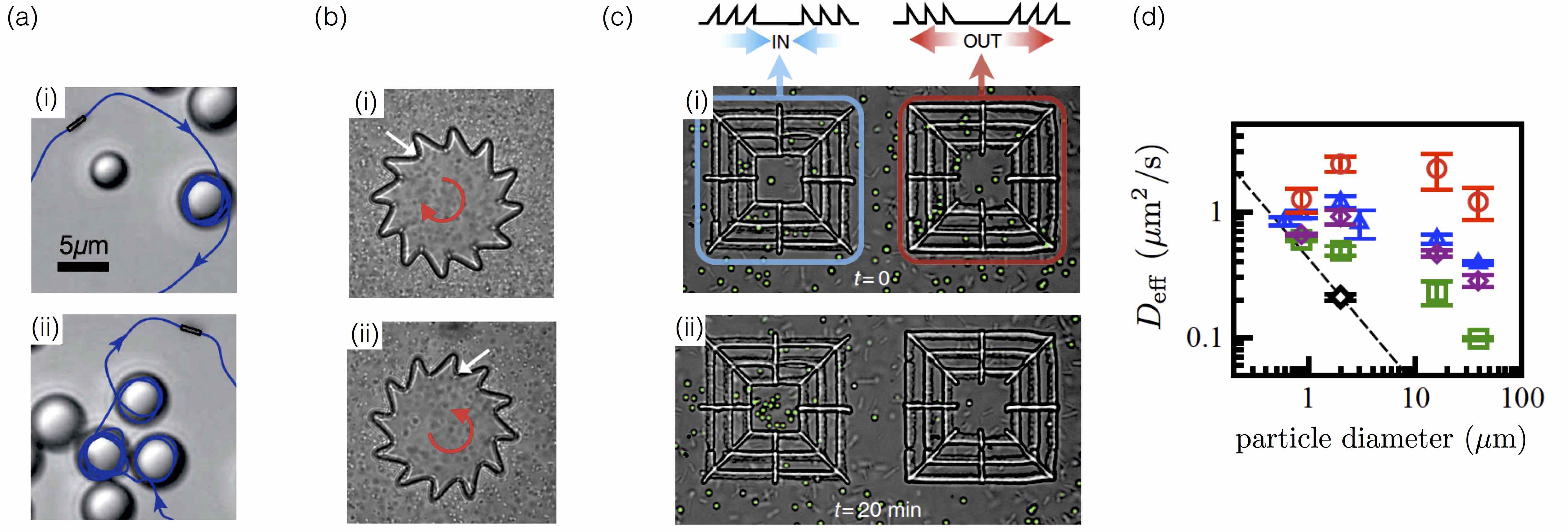}
\caption{Interplay between passive and active particles. (a) Passive spheres temporarily capture micro-swimmers. The active colloids are Au-Pt  rods moving in aqueous hydrogen peroxide \cite{Takagi2014}. Trajectories of single rods are shown in blue. (b) Bacterially driven microgears: Collisions between swimming bacteria and gears drive clockwise or counterclockwise rotation depending on the orientation of the teeth (i vs ii) \cite{Wong2013}. (c) Surface topology in the presence of motile bacteria guides an (i) initial distribution of colloids to (ii) either aggregated (left) or depleted (right) regions \cite{Leonardo2013}. (d) Tracers moving in bacterial baths exhibit anomolous size and concentration dependent  
effective particle diffusivities $D_{\mathrm{eff}}$. The dashed line is the particle thermal diffusivity given by the classical Stokes-Einstein relationship. Colors correspond to bacterial concentrations (lowest, green  to highest, red) \cite{Patteson2015_2}.}
\end{figure*}

 \subsubsection{Surfaces and swimming}

{Swimming near surfaces plays a central role in many important biological settings. For example, sperm cells are guided by microgrooves and fluid flows along the surface of the female reproductive path \cite{Wu2015}. Similarly, bacteria swim and adhere to surfaces before biofilms can form \cite{Stone2015}. Hydrodynamic effects typically drive the organisms towards solid surfaces \cite{Lauga2009}. Once in the vicinity of such boundaries, these swimmers exhibit additional fascinating features not observed in bulk swimming. Features include circular trajectories \cite{lauga2006}, rheotaxis \cite{Rothschild1961,Marcos2012}, and suppressed bacterial tumbling \cite{Stocker2014_2}; these arise specifically due to the no-slip boundary condition near the solid interface. A bacterium swimming along the surface experiences higher shear stress compared to the bulk. Combined with the rotations of the cell body and the flagellar bundle, the cell swims in a clockwise trajectory as viewed from above the surface. Interestingly, at an air-water interface where the boundary condition is full-slip, the effect is opposite i.e., the bacterium swims counter-clockwise. Applying an external shearing flow biases the trajectory by reorienting the rod-shaped bacteria - the organisms then tend to swim upstream, a feature known as rheotaxis.}

{Furthermore, swimming near surfaces can be influenced by the fact that the ambient medium is often a non-Newtonian fluid. Recent theoretical work \cite{Ardekani2014} has shown that residence times of a bacterium near a solid wall can be significantly enhanced due to fluid elasticity. Also, at an air-fluid surface, organic molecules in nutrient-rich growth media can adsorb at the interface creating a local, high-viscosity region. This modifies the full-slip boundary condition, and the interface acts more like a solid boundary \cite{Tang2013,Lauga2014_2}. Even in the case of clean air-water surface, the possibility of interface deformations introduces interesting characteristics. For example, nonlinearities due surface tension can even propel bodies that have reciprocal swimming strokes \cite{Lauga2008}, a strategy that does not work at hard surfaces or in the bulk of Newtonian fluids for low-Reynolds number swimmers.}

{We have discussed slip and boundary effects near surfaces external to the swimming active colloids. An interesting question that arises is the role of boundary conditions at the surface of the swimmer. Indeed, in microstructured fluids, the swimmer might modify the local microstructure and {\em slip} through the medium.  In polymer solutions, shear-thinning viscosity effects {\it E. coli} may contribute to the increase in bacterial cell velocity \cite{Wilson2014}. For the undulatory swimmer \textit{C. elegans}, shear-thinning viscosities in polymer solutions have little to no effect on the swimming speed \cite{Gagnon2014_2}, while in colloidal suspensions, the swimming speed increases \cite{Weitz2015}. More work is needed to fully understand the role of slip at the surface of the swimmer.}

 \subsection{Synthetic swimmers in complex fluids}

As shown above, fluid rheology can significantly affect the swimming behavior of microorganisms. In this section, we will explore a different, and perhaps simpler, question: Can the fluid non-Newtonian rheological properties \textit{enable} propulsion? To answer this question, one needs to think back to the ``scallop theorem" which tell us that only non-reciprocal deformations of the swimmer can break time reversal symmetry and result in net motion. One of the main assumptions of the theorem is that the swimmer is moving in a purely viscous (Newtonian) fluid. If we relax this assumption, then it may be possible to break kinematic reversibility and achieve net motion even for reciprocal swimming strokes. A simple thought experiment may lead to this conclusion: Consider Purcell's scallop now swimming in a shear-thinning fluid. One can imagine that if the scallop opens and closes its mouth at different rates, it may encounter different shear viscosities and viscous stresses as it opens and closes its mouth; the shear-rates associate with opening and closing motions are different. This viscous stress ``imbalance" may be enough to lead to net motion. It is important to note, however, that just because kinematic reversibility is broken, it does not mean that one has achieved efficient propulsion; it only means that propulsion is possible. 

This possibility  - that the fluid non-Newtonian rheological properties can enable propulsion  - has been explored theoretically for a handful of special cases, particularly for viscoelastic fluids. Normand and Lauga \cite{Normand2008} and Pak, Normand, and Lauga \cite{Pak2010} analyzed the role of fluid elasticity on model tethered-flappers, which execute prototypical small amplitude reciprocal motion. Analysis of the flow fields generated by swimmers moving in Oldroyd-B and FENE-P model fluids suggests that elastic effects can generate forces that scale quadratically with the amplitude of the motion. This demonstrates that normal-stress differences due to elasticity can be exploited to enable propulsive forces, circumventing the scallop theorem. In a related study, Fu, Wolgemuth and Powers \cite{Fu2009} studied a variation of the Taylor sheet problem \cite{Taylor1951}: an infinitely long ``wiggling" cylinder in an Oldroyd-B fluid. Using domain perturbation techniques, the authors find that net motion is possible for reciprocal motion in which the backward and forward strokes occur at different rates, possible due to an imbalance of generated normal stress differences. Another case is a `squirming' sphere, which executes small-amplitude motion along its surface. Using concepts drawn from differential geometry and utilizing the reciprocal theorem, Lauga \cite{Lauga2009_2} found that in an Oldroyd-B fluid the accumulation of local elastic stresses drives the sphere forward. This is true even when the surface motion is time-reversible, breaking free of the constraints imposed by the scallop theorem. These theoretical works clearly allude to the possibility of elasticity-enabled propulsion.

Perhaps the first experimental demonstration of elasticity-enabled propulsion was provided by Arratia and co-workers \cite{Keim2012,Gagnon2014}. In these experiments, an asymmetric particle (in this case a dimer) is actuated by an external magnetic field and it is forced to execute periodic reciprocal strokes, which results in no net motion in viscous Newtonian fluids. In a dilute polymeric solution \cite{Keim2012}, however, net motion is achieved by elastic stresses which do not entirely cancel out over one forcing period,  but instead have a small rectified component that accumulates. This elasticity-enabled propulsion has also been observed in ``structured" fluids such as worm-like micellar solutions \cite{Gagnon2014}. Furthermore, propulsion may be enabled by other non-Newtonian fluid properties, such as shear-rate dependent viscosity. Indeed, Qiu \emph{et al.}  \cite{Peer2014} have observed the propulsion of reciprocally-swimming micro-scallops in shear-thinning and shear-thickening fluids.

{Promising advances have been in the design of self-propelling micro- and nano-swimmers. These advances include using hydrogen peroxide as a fuel source \cite{Mallouck2009} and electric field-induced polarization \cite{Kuhn2014}. However, powering self-propelling synthetic swimmers still remains a challenge \cite{Howse2010}. Detailed experiments \cite{Keim2012,Gagnon2014} demonstrate that driven artificial swimmers can move through complex fluids with only reciprocal actuations and a simple body shape. These synthetic externally-driven swimmers are appealing for biological applications since their propulsive mechanism is less complicated than alternate strategies. The propulsive efficiency of micro-swimmers is commonly defined as the ratio of the power to drag the swimmer at its period-average velocity to the total power dissipated by the fluid during a period. The synthetic reciprocal swimmers in polymeric solutions have propulsive efficiencies ($\approx$1$\%$ \cite{Keim2012}) similar to those of non-reciprocal swimmers in Newtonian fluids, including magnetic torque-driven helical micro-robots ($\approx 1\%$ \cite{Peyer2013}) and self-propelling force-free bacteria ($\approx 2\%$ \cite{Wu2006}).} Further understanding of factors controlling this efficiency could greatly simplify fabrication of micro-swimmers.

\section{Suspensions of Active Colloids \& Swimmers}

In general, a single active entity in a fluid - Newtonian or complex - behaves very differently from a suspension comprised of multiple such entities. Examples are shown in { Fig. 5}. Interactions between multiple swimmers (or active colloids) can lead to many fascinating phenomena not seen in suspension of passive particles at equilibrium including anomalous density and velocity fluctuations, large scale vortices and jets, and traveling bands and localized asters. Identifying means to relate the microstructural features (e.g. swimmer local orientation) to macrostructural properties and bulk phenomena would yield ways to control, manipulate, and even direct the properties in these novel living systems.

 \subsection{Dilute suspensions of active particles}
 
A suspension of active colloids is considered dilute when interactions among particles are negligible. Even in the absence of particle interactions, however, the interplay between activity and the fluidic environment, as reviewed below, leads to novel and even unexpected phenomena.

\subsubsection{Active particles in Newtonian fluids}

In the absence of activity, the shear viscosity $\eta$ of a dilute suspensions of (passive) hard spheres is given by the Einstein relation, $\eta = \eta_s(1+{5 \over2} \phi)$ \cite{Einstein1906}, where $\eta_s$ is the viscosity of the suspending fluid and $\phi$ is the  volume fraction of the particles. In the presence of activity, however, the shear viscosity can be a strong function of the microorganisms swimming kinematics. We will briefly discuss the origins of this behavior below.

By using a kinetic theory based approach and solving the Fokker-Plank equation for the distribution of particle orientations under shear, Saintillan \cite{Saintillan2010} showed that for a dilute suspension of force dipoles, the zero-shear viscosity $\eta$ still follows an Einstein-like relation, $\eta=\eta_s (1+K\phi)$, with the constant $K$ now related to swimmer kinematics. For pushers, $K<0$, while for pullers, $K>0$. This leads to an interesting result: activity can either enhance or reduce the fluid viscosity depending on the swimmer kinematics (puller or pusher). 

Due to the exceptionally low shear rates and stresses needed to realize these potential modifications in viscosity, experimental verifications of theories have been limited \cite{Sokolov2009_2,Rafai2010,Clement2015}. In 2009, Sokolov and Aranson \cite{Sokolov2009_2} presented some of the first experimental evidence of activity-modified viscosity in a fluid film of pushers ({\it Bacillus subtilis}). They found that the presence of bacteria significantly reduces the suspension effective viscosity. Subsequent experiments using shear rheometers have shown that the fluid viscosity can be effectively larger in suspensions of {\it C. reinhardtii} (pullers)~\cite{Rafai2010} or lower in suspensions of {\it E. coli} (pushers)~\cite{Clement2015} compared to the case of passive particles (non-motile organisms) for the same shear-rates. Activity also seems to affect the suspension extensional viscosity in a similar way \cite{McDonnell2015,Saintillan2010_2}.
  
Clearly, activity has a fascinating effect on the viscosity of active suspensions; theoretical and numerical investigations seem to predict a regime in which the viscosity of the suspension can be lower than the viscosity of the suspending fluid. This striking phenomenon has been recently observed in experiments for {\it E. coli} (pushers)~\cite{Clement2015}, where it was also found that the suspension viscosity linearly decreased as the bacterial concentration increased (in the dilute regime). Despite such advances, however, it has been a challenge to experimentally visualize the evolution of the microstructure (particle positions and orientations) during the rheological (viscosity) measurements. This type of information and measurements are critical to obtain insights into the physical mechanisms leading to this ``vanishing" viscosity phenomenon in bacterial suspensions. 

 \subsubsection{Active particles in complex fluids}

Given the evidence that bacterial activity can alter suspension viscosity, it is natural to expect that the interaction of active particles or microorganisms with the fluid microstructure (polymers, particles, liquid crystals, cells, and networks) in complex fluids can also lead to interesting phenomena. Indeed, an extreme example of this is how even dilute concentrations of bacteria can disrupt long range order in lyotropic liquid crystals. The presence of bacteria can locally melt the underlying nematic order and generate large scale undulations with a length scale that balances bacterial activity and the anisotropic viscoelasticity of the suspending liquid crystal ~\cite{Aranson2013_2}. Related experiments using active liquid crystals comprised of reconstituted microtubule-motor mixtures \cite{Decamp2015} suggest similar disruptive effects on long range order. In this case, the active entities are motile defects, which generate flow, and are spontaneously created and annihilated within the ambient environment, as shown in { Fig. 5e}. These studies illustrate how even dilute concentrations of active particles can locally deform and activate the microstructure of complex fluids. These synergistic and dynamic materials possess qualities (new temporal and spatial scales) distinct from both passive complex fluids and suspensions of active particles in Newtonian fluids. 

The microstructure of complex fluids, however, does not simply submit to the flow generated by active particles. Instead, as discussed in Section 2.2 and 2.3, the microstructure couples to the active particles, altering their swimming gait and speed. Indeed, the microstructure can even be exploited to adaptively guide active particles. For instance, it has recently been shown that the underlying nematic structure of lyotropic liquid crystals can align bacteria, controlling their motility and direction \cite{Aranson2013_2}. The nematic director can even set the bacterial direction near walls, where near-wall hydrodynamic torques can reorient cells \cite{Abbott2015}. In recent experiments, Trivedi {\it et al.} \cite{Trevedi2015} demonstrated that in lyotropic liquid crystals bacteria can transport particles and non-motile eukaryote cells along the nematic director. Conversely, passive particles ($\approx$ 1-15 $\mu$m diameter) can be used to manipulate and capture active particles (self-propelled Au-Pt rods) \cite{Takagi2014}, which tend to orbit along surfaces of passive particles, as shown in { Fig. 6a}. Together, these works seem to mirror the trafficking of cargo in cells by active motors \cite{Howard2001} and suggest novel methods to transport active and passive components of these living, complex fluids, some of which are highlighted in { Fig. 6}.

\subsection{Non-dilute suspensions of active particles}

The investigations briefly discussed above highlight the striking role of activity on material fluid properties even in the dilute regime when particle interactions are negligible. As the concentration of particles increases, however, the particle interactions (either steric and aligning interactions or hydrodynamic interactions) can suddenly give rise to collective motion. Interestingly, as reviewed next (Section 3.2.1), even when the suspending medium is not a complex fluid, the collective dynamics of the active particles can lead to non-Newtonian suspension properties. The interplay between collective particle dynamics and a suspending complex fluid are yet to be explored in detail, with the exceptions of the works mentioned in Sec. 3.2.2.

\subsubsection{Active particles in Newtonian fluids}

Perhaps one of the first models for collective motion in the absence of fluid hydrodynamic interactions was proposed by Toner and Tu \cite{Toner1998} using a modification of the classical liquid crystal model. This seminal work has been significantly extended theoretically to cover a range of interactions. Interestingly, these models as well as simpler discrete agent-based simulations are able to capture many of the universals features observed in natural active colloidal systems including flocking and collective behavior.

In order to incorporate the role of fluid interactions, recent mean-field models use dipole approximations in simple (Newtonian) fluids.  Recent reviews by Koch and Subramanian ~\cite{Koch2011} and Marchetti, {\it et al.} \cite{Marchetti2013} summarize linear stability analyses of these mean-field models.  The general consensus of these studies is that hydrodynamic interactions mediated by the fluid can, in some cases, destabilize homogeneous suspensions and assist collectively moving states.

As mentioned above, even when the suspending fluid is Newtonian, interactions between active particles can induce non-Newtonian features, such as elasticity. In order to model the rheology of active suspensions, Hatwalne {\it et al.} \cite{Hatwalne2004} generalized the kinetic equations for liquid crystals and obtained a general expression for frequency-dependent stress in an oscillatory shear flow. This stress depends on the detailed swimming kinematics (pusher or puller), the active correlation times, and the density. Importantly, the theory predicts that -- as the system approaches an orientational-order transition -- this previously Newtonian fluid begins to exhibit elasticity, with elastic stresses than increase with the orientational order. This work highlights how the active particle microstructure can dramatically alter the bulk material properties.

\subsubsection{Active particles in complex fluids}
The implications of a suspending non-Newtonian fluid on the bulk material properties of active fluids, such as collective behavior, are only now starting to be explored. In this regard, a series of recent theoretical studies by Bozorgi and Underhill \cite{Underhill2011, Underhill2013, Underhill2014} reveal how features unique to non-Newtonian fluids, i. e. fluid relaxation times, may result in coordinated collective motion of active colloids. In this sequence of papers, the authors analyze the evolution of an initially uniform distribution of interacting axisymmetric colloids moving in three canonical viscoelastic fluids, namely Oldroyd-B, Maxwell, and  generalized viscoelastic fluid models.  Linear stability analysis reveals that the emergence of instabilities and collective motion is controlled by a number of parameters: colloidal dynamics (such as translational and rotational diffusivities), fluid material properties (such as relaxation times and viscosity) and interaction-specific features (such as the time for hydrodynamic interactions to reorient colloids). The authors find that increasing particle translational diffusivity always hinders the onset of collective motion; whereas, rotational diffusivity always hinders instabilities when the suspending fluid is Newtonian. Surprisingly, when the suspending fluid is non-Newtonian, rotational diffusivity can promote the emergence of these collectively moving states \cite{Underhill2013}. Recent experiments \cite{Patteson2015} have shown that fluid viscosity and elasticity impacts the translational and rotational diffusivity of individual, non-interacting {\it E. coli}. This suggests that the presence or absence of collective motion in active fluids can be controlled by simply tuning the mechanical properties of the suspending medium.

The viscoelasticity of the suspending fluid not only affects the emergence of collective motion but also influences the pseudo-steady flows that are eventually established after the onset. Fluid elasticity can force strongly collective states to transition to weakly collective ("suppressed") states with less long-range structure. Indeed, a recent theoretical study \cite{Underhill2014} predicts that the self-driven structures that emerge when active colloids move in complex fluids have considerably different length and time scales from the Newtonian analogue.

\section{Future and Perspectives}

True appreciation of a complex mosaic is gleaned only by examining each individual component and appreciating its role in the way it fits into, integrally contributes to, and thus {\em forms} the whole. Similarly, a theoretical and practical knowledge of the dynamics of active colloids in complex fluids necessitates understanding at all scales. It is clear from a review of the literature that significant progress has been made in this regard. Still, many aspects remain to be uncovered and understood. Crucial to this process is the connection between theory, simulations, and experiments in a meaningful way. We conclude by focusing on three over-arching questions and aspects that are yet to be resolved satisfactorily: (i) passive versus active response of individual microorganisms, (ii) complex interactions among particles, and (iii) constitutive equations for suspensions of active colloids.

\subsection{Passive and active response}

The first step in understanding active fluids is to understand swimmer-fluid interactions at the level of a single swimmer. As is evident from a review of the literature of single swimmers in complex fluids (Section 2), the fluid microstructure and swimming kinematics together impact the organism's motility in a non-linear manner. Changes in the swimming behavior, i.e. the swimming gait, of a living organism may however not be a purely passive response. Microorganisms can use mechanosensation to actively control their swimming modes and attempt to adapt to a (complex) fluidic environment. For instance, it is known that sperm can exhibit different beating shapes in response to the properties of the fluid through which they swim \cite{Woolley2001}. In reality, the interplay among the fluid stresses, the passive material properties of the swimmer's body, and the active forces produced by the internal motors produces the configurational changes that make up the swimming gait \cite{Guy2014, Fauci2006}. Separating the active swimmer response aspects, which may have behavioral and adaptive evolutionary, to the ambient fluid from its passive response is very challenging and yet has promising implications for the design of synthetic swimmers, micro-robots, and overall understanding of swimming. 

A natural step is to develop models that allow for independent tuning of these three complementary responses that together affect the swimmer's response. Experiments that will provide insight in this regard must simultaneously visualize the single swimmers and the time-dependent surrounding flow as well as obtain information on the organism's ``activity." For the case of the nematode \textit{C. elegans}, this would mean visualizing calcium ions in the muscle cells during swimming, for example. Calcium imaging has potential for investigating the molecular mechanisms involved in muscular contraction and can even be used to infer neural activity. Due to the wealth of genetic information as well as well-developed electrophysiology and optical techniques (e.g. opto-genetics), the nematode \textit{C. elegans} is a promising organism for such investigations.

\subsection{Particle interactions}

A second challenge  is in understanding and manipulating local particle interactions between active colloids in order to control and direct bulk collective behavior. Although theories and simulations have had some success in linking simple particle interactions (steric, aligning, and hydrodynamic) with the emergence and dynamics of collective motion \cite{Dreyfus2005,Hagan2013}, in general, active colloidal particles interact in a plethora of ways. Examples include electro-static repulsions, short-range attractions \cite{Wilson2012} and adhesion \cite{Lisa2015}. In experimental systems, these interactions are difficult to distinguish, seperate, and even quantify. For instance, quorum sensing is used by bacteria to actively change their behavior at high bacterial densities \cite{Stocker2014}. However, separating this chemically based signaling response from hydrodynamical based interactions is difficult. These challenges highlight the need for simple (nonliving) experimental systems, such as self-propelled phoretic particles \cite{Bartolo2013,Palacci2013} and externally actuated particles \cite{Wang2014,Koser2013_2}, for independently varying particle interactions and testing theories on active complex fluids.

Furthermore, most simulations and theories focus on monodisperse active particles that behave identically. However, diversity is a fundamental aspect of life and is reflected in variations in the kinematics between individual microorganisms of a species. Polydispersivity due to extraneous suspended material such as particles, polymers or other motile or non-motile cells is inherently seen in many natural environments. Simulations \cite{Cates2015} have shown that bidisperse (motile and non-motile) particles phase-separate under appropriate conditions. Phase separation has also been observed in experiments on bacteria moving in polymeric solutions  ({ Fig. 5d}) \cite{Wilson2012}. Recent investigations have also shown that hydrodynamic interactions between bacteria and passive spheres can enable larger particles to diffuse more rapidly than smaller ones ~\cite{Koch2014,Patteson2015_2}, as shown in { Fig. 6d}. These studies highlight fascinating phenomena that occurs when passive and active particles interact; topics that are fertile ground for vigorous exploration.

The motion of passive particles or tracers can additionally be used to probe the characteristics of the environment such as the local viscosity, permeability and temperature. This, in fact, forms the basis of microrheology in passive systems - both in and out of equilibrium ~\cite{Squires2009,Crocker2000}. These powerful and established techniques have recently been extended to active materials and have shed light on spatial and temporal inhomogeneities in these systems, an example being bacterial suspensions \cite{Chen2007}. Complementarily, the dynamics of soft and filamentous particles can yield additional information not obtainable using rigid tracers: fluorescently stained DNA molecules can be used to estimate local, history-dependent stresses in fluid flow  \cite{Steinberg2010,Chu1999,Doyle2000}. Simulations suggest that the collective dynamics of active particles leads to unusual swelling of polymer molecules \cite{Andreas2014}, whereas we have shown experimentally that the flow generated by individual motile {\it E. coli} are sufficiently strong to stretch DNA polymer molecules~\cite{Patteson2015}. These studies suggest how particles, hard and soft, can be used to gauge and understand activity.

\subsection{Constitutive Equations}

As reviewed above, the combination of active and passive ingredients within an fluid can lead to fascinating and sometimes non-intuitive features. Accurate and predictive models for these features will need to connect microscale physics to macroscale structures. Classically, as in the case of passive albeit complex fluids, the bulk response at large hydrodynamic scales is described through the use of constitutive equations describing the features of the fluid, such as the stress and pressure, in terms of variables such as the deformation and deformation rate. These constitutive equations are related to the microstructure of the complex fluid using concepts from statistical physics and kinetic theory. For example, the bulk polymeric stress in a flowing polymer suspension can be related to the ensemble averaged mean stretch of  the polymers provided polymer-polymer and polymer-solvent interactions are known. There have been many focused efforts to extend these techniques to active complex fluids. We highlight two aspects of this broadly-defined problem related to thermodynamic and mechanical properties of active fluids.

First, can thermodynamic variables, such as temperature and chemical potential, be used in a meaningful way to characterize active suspensions \cite{Wilson2012,Palacci2010,Gopinath2011,Brady2015_2}? Although the notion of temperature proves useful in some contexts \cite{Palacci2010,Brady2015_2}, in general, studies \cite{Wilson2012} suggest that the temperature cannot be defined unambiguously for active systems. The second aspect relates to whether mechanical properties such as the stress tensor and pressure can be defined for active fluids and whether these properties are consistent when determined in independent ways. Closely related to this issue is if and when equations of states for active colloidal fluids can be written down. As an example, for passive fluids in thermal equilibrium, pressure may be defined as (i) either the force per unit area exerted by the fluid on its containing vessel, (ii) through an equation of state that relates pressure to bulk thermodynamic properties such as density and temperature and (iii) via hydrodynamic principles, as the trace of a macroscopic stress tensor. In thermal equilibrium, all these definitions of pressure coincide but far from equilibrium, these definitions may not. Theories and simulations are beginning to explore when these definitions may converge for active fluids.

Recent models for the macroscopic dynamics of active colloids in Newtonian fluids have introduced the expressions for the bulk stress tensor (thereby incorporating activity) and defined pressure as its trace \cite{Marchetti2013,Brady2015}. How this pressure relates to the mechanical definition of pressure -- as the force per unit area acting on a bounding surface -- is an open question. Solon {\it et al.} \cite{Solon2015} have shown that, when momentum coupling to the ambient fluid is negligible, the pressure exerted on a wall by an assembly of self-propelled colloids depends on the microscopic details of colloid-wall interactions. This remarkable result strongly suggests that generic equations of state for active colloidal systems cannot be written down. However even when an equation of state cannot be written down, the notion of pressure remains useful \cite{Brady2015,Solon2015}.

We emphasize that it is unclear how these results would change for a real active colloidal complex fluid. In physical systems, there is full momentum conservation (for both particles and suspending fluid) in the bulk as well as inter-colloidal interactions mediated by the fluid. The general case that involves full hydrodynamics -- be they Newtonian or non-Newtonian -- is still unexplored and offers exciting challenges to both theorists as well as experimentalists. We anticipate that these studies will provide a foundation for bridging the gap between inorganic complex fluids and organic active matter. After all, as Rinne wrote even as early as 1930, ``this gap does not exist, since the sperms, which are undoubtedly living, are at the same time liquid crystals \cite{Rinne1930}.''

\section*{Acknowledgments}

We acknowledge funding from NSF-CBET-1437482 and NSF DMR-1104705.


\section*{References$^{*}$\footnote{* - of special interest}}




\end{document}